\documentclass[aps,prc,reprint,showpacs,showkeys,floatfix]{revtex4-1}

\usepackage[utf8]{inputenc}
\usepackage{graphicx}
\usepackage{bm}
\usepackage{amsmath}
\usepackage{amssymb}
\usepackage[colorlinks]{hyperref}
\usepackage{xspace}
\usepackage{upgreek}
\usepackage{siunitx}
\usepackage[capitalize]{cleveref}
\crefname{figure}{FIG.}{FIGS.}
\Crefname{figure}{Figure}{Figures}

\renewcommand{\vec}[1]{\mathbf{#1}}
\newcommand{\premarker}{\ensuremath{^\text{pre}}}
\newcommand{\postmarker}{\ensuremath{^\text{post}}}
\newcommand{\mass}{\ensuremath{M}\xspace}
\newcommand{\energy}{\ensuremath{E}\xspace}
\newcommand{\mpre}{\ensuremath{\mass\premarker}\xspace}
\newcommand{\mprem}{\ensuremath{\mass^\text{m}}\xspace}
\newcommand{\mpret}{\ensuremath{\mass^\text{c}}\xspace}
\newcommand{\mpost}{\ensuremath{\mass\postmarker}\xspace}
\newcommand{\mn}{\ensuremath{m_\text{n}}\xspace}
\newcommand{\en}{\ensuremath{\energy_\text{n}}\xspace}
\newcommand{\msum}{\ensuremath{\mass_\text{sum}}\xspace}
\newcommand{\mcn}{\ensuremath{\mass_\text{Cf-252}}\xspace}
\newcommand{\epre}{\ensuremath{\energy\premarker}\xspace}
\newcommand{\epost}{\ensuremath{\energy\postmarker}\xspace}
\newcommand{\tke}{\text{TKE}\xspace}
\newcommand{\tkepre}{\ensuremath{\tke\premarker}\xspace}

\newcommand{\vel}{\ensuremath{v}\xspace}
\newcommand{\velpre}{\ensuremath{\vel\premarker}\xspace}
\newcommand{\velpost}{\ensuremath{\vel\postmarker}\xspace}

\newcommand{\twoe}{\texorpdfstring{$2\energy$}{2E}\xspace}
\newcommand{\twoetwov}{\texorpdfstring{$2\energy$--$2\vel$}{2E-2v}\xspace}
\newcommand{\nubar}{\ensuremath{\bar\nu}\xspace}
\newcommand{\nubarm}{\ensuremath{\bar\nu^\text{m}}\xspace}
\newcommand{\nubart}{\ensuremath{\bar\nu^\text{c}}\xspace}

\begin{document}

\title{Defective fission correlation data from the 2E-2v method}
\date{\today}
\author{Kaj Jansson}
\email{kaj.jansson@physics.uu.se}
\author{Ali Al-Adili}
\email{ali.al-adili@physics.uu.se}
\thanks{Corresponding author}
\author{Erik \surname{Andersson Sund\'en}}
\email{erik.andersson-sunden@physics.uu.se}
\author{Stephan Pomp}
\email{stephan.pomp@physics.uu.se}
\affiliation{Department of Physics and Astronomy, Uppsala University, Box~516, 751~20 Uppsala, Sweden}
\author{Alf Göök}
\email{alf.gook@ec.europa.eu}
\author{Stephan Oberstedt}
\email{stephan.oberstedt@ec.europa.eu}
\affiliation{European Commission, DG Joint Research Centre, Directorate G - Nuclear Safety and Security, Unit G.2 SN3S, 2440 Geel, Belgium}

\begin{abstract}
The double-energy double-velocity ($2E$-$2v$) method allows assessing fission-fragment mass yields prior to and after prompt neutron emission with high resolution. It is, therefore, considered as a complementary technique to assess average prompt neutron multiplicity as a function of fragment properties. We have studied the intrinsic features of the $2E$-$2v$ method by means of event-wise generated fission-fragment data and found severe short-comings in the method itself as well as in some common practice of application. We find that the $2E$-$2v$ method leads to large deviations in the correlation between the prompt neutron multiplicity and pre-neutron mass, which deforms and exaggerates the so called `sawtooth' shape of $\bar{\nu}(A)$. We have identified the treatment of prompt neutron emission from the fragments as the origin of the problem. The intrinsic nature of this deficiency, risk to render $2E$-$2v$ experiments much less interesting. We suggest a method to correct the $2E$-$2v$ data, and recommend applying this method to previous data acquired in $2E$-$2v$ experiments, as well.
\end{abstract}



\maketitle

Fission-fragment (FF) spectroscopy has been carried out since the very beginning of the discovery of nuclear fission. It represents an important gateway for understanding the dynamics of fission. In particular the number of prompt neutrons emitted from each highly excited FF and its correlation with FF properties (mass and total kinetic energy) is a key to better understand how the excitation energy is shared between the fragments. Together with other observables, e.g., isomeric yield ratios and prompt $\upgamma$-rays, they may provide a deeper insight into the dynamics of the fission process. High-resolution fragment yields may reveal the underlying nuclear structure, observed, e.g., through the so-called odd-even staggering in the observed mass yields.

One principal technique for investigating FF properties, e.g., pre-neutron mass yields and total kinetic energy (\tke) distributions, is based on the measurement of both fragment kinetic energies; the so-called double-energy (\twoe) method \cite{2e}. The conducted \twoe studies are numerous, but rarely with a pre-neutron mass resolution no better than \SI{4}{\amu}.

Two techniques bringing major improvements to the mass resolution are the energy-velocity method, exploited at the LOHENGRIN recoil mass separator \cite{lohengrin}, and the double-energy double-velocity method (\twoetwov), which was introduced via the concept of COSI-FAN-TUTTE \cite{oed}. The \twoetwov method is currently under development world-wide, e.g., VERDI \cite{verdi}, SPIDER \cite{spider} (and future MegaSPIDER), FALSTAFF \cite{falstaff} and STEFF \cite{steff}. All the present realizations of the \twoetwov method promise a resolution of the pre-neutron mass better than \SI{2}{\amu}.

While the \twoe method relies on prompt neutron emission data as an input to the data analysis, the \twoetwov technique promises provision of those data from the independent analysis of the experimental data. A focus is put on prompt-neutron FF correlations, which merit more and more attention by theoreticians and modelists. As many experiments have used a \twoe analysis for calibration purposes, we studied aspects of this technique too. We will demonstrate inherent short-comings of both the techniques and how they affect the \twoetwov data. We also suggest a way to remedy both the \twoetwov method and previously obtained \twoetwov data.

To facilitate a clean test of the analysis methods, they are not tested with experimental data. Instead we generated realistic synthetic data, in order to know the true values of all variables exactly. The generated data do not need to replicate reality perfectly since we merely want to know how well the analysis method will reproduce the synthetic data. However, it is beneficent to work with reasonably realistic data.

The \textsc{gef} code \cite{gef} was used to generate \num{1e6} fission events of the studied fissioning system. The fragment recoil upon neutron emission is not correctly handled by \textsc{gef}, so the post neutron state was recalculated based on the pre-neutron state as well as a list of \emph{Center-of-Momentum} (CoM) energies of the emitted neutrons for both fragments. In both cases the information was supplied by \textsc{gef}. For each neutron energy in the list, a random direction was selected isotropically in the fragment's CoM frame, and the emission kinematics was calculated in full.

Since the \textsc{gef} output only contained the atomic and mass numbers, the fragment masses were looked up in \emph{The Ame2012 atomic mass evaluation} \cite{mass12-1,*mass12-2}. The processed events were written to a \textsc{root} file \cite{root}, later to be analyzed by the \twoetwov method, as well as the \twoe method, both described below.

In our `ideal' study of the \twoetwov method no resolution or other detector effects are included. In the analysis relativistic kinematics are used, but for simplicity, we show the classical equivalent expressions for how the relevant quantities are derived:
\begin{align}
\label{eq:mpre2e2v}\mpre_{1,2} &= \msum \frac{\velpre_{2,1}}{\velpre_1+\velpre_2} \\
\label{eq:mpost2e2v}\mpost_{1,2} &= \frac{2\epost_{1,2}}{(\velpost_{1,2}{})^2} \\
\label{eq:epre2e2v}\epre_{1,2} &= \frac{\mpre_{1,2}(\velpre_{1,2}{})^2}{2} \\
\label{eq:nu2e2v}\nu_{1,2} &= \frac{\mpre_{1,2}-\mpost_{1,2}}{\mn},
\end{align}
where $\msum = \mpre_1+\mpre_2 = \mcn - \tkepre / c^2$, \mass denotes mass, \vel velocity, \energy energy and $\nu$ neutron multiplicity. The momentum of the incoming particle has been neglected since we have only studied spontaneous fission or fission induced by thermal neutrons.

The \twoetwov method makes the assumption of isotropic neutron emission in the CoM frame, which \emph{almost} leads to a conservation of the average values of the velocities as pointed out by Stein~\cite{stein}. It can be shown that the difference is less than \SI{0.01}{\percent}. The velocities are not conserved event-wise, but since \velpre is not measured, the \twoetwov method approximates \velpre by \velpost in \cref{eq:mpre2e2v,eq:epre2e2v}. The error of this estimation of \velpre is, in the $^{252}$Cf(\textit{sf}) case, approximately equivalent to a \SI{1}{\percent} normally distributed random error in \velpre. 

Contrary to the \twoetwov method, the \twoe method requires additional information since only the two energies are measured.  The \twoe analysis needs the neutron multiplicity $\nu$ to be parametrized based on information acquired from previous measurements or model calculations. In our case, the information comes from \textsc{gef} rather than experimental studies.

For each event, the \twoe method iterates \cref{eq:mpost2e,eq:epre2e,eq:mpre2e} until the masses (and energies) converge.
\begin{align}
\label{eq:mpost2e} \mpost_{1,2}(i+1) &= \mpre_{1,2}(i) - \nu_{1,2}\,\mn \\
\label{eq:epre2e} \epre_{1,2}(i+1) &= \epost_{1,2} \frac{\mpre_{1,2}(i)}{\mpost_{1,2}(i+1)} \\
\label{eq:mpre2e} \mpre_{1,2}(i+1) &= \msum \frac{\epre_{2,1}(i+1)}{\epre_1(i+1)+\epre_2(i+1)}
\end{align}
When all masses changed by less than \SI{1e-10}{\amu} between two subsequent iterations, convergence was considered reached. Typically, only a few iterations were needed.

In \cref{eq:epre2e}, two terms for each emitted neutron have been left out:
\[
\frac{\mn}{\mpost} \en^\text{CoM}\quad \text{and} \quad \mn\vec{\velpre}\vec{v}_\text{n}^\text{CoM}. 
\]
The first term is in the order of tens of \si{\keV}, while the second term vanishes on average, assuming a isotropic neutron emission in the CoM frame of the fragment. Just as in the \twoetwov case, this approximation is only strictly valid for average quantities.

\begin{figure}[tb]
\centering
\includegraphics[width=0.5\textwidth]{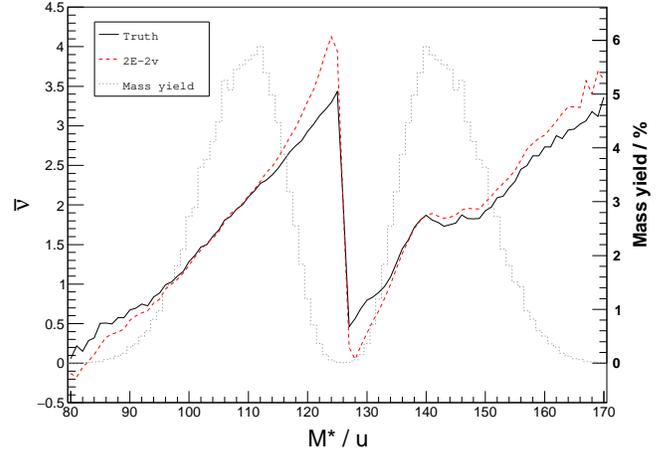}
\caption{\label{fig:nubar}(Color online) The \twoetwov method shows discrepancies, which can be traced back to the neutron emission. The `true' \nubar and \mpre distributions are shown as references. }
\end{figure}

Both the \twoetwov and the \twoe method have been found to, on average, reconstruct the proper mass within about \SI{0.1}{\amu}. A slight overestimation of \mpre is expected due to the excitation energy of the FFs, which is unaccounted for in \cref{eq:mpre2e2v,eq:mpre2e}. Although both methods reproduce the one-dimensional mass and energy spectra well, discrepancies start to show as one looks at other derived quantities. In \cref{fig:nubar} the resulting $\nubar(\mpre)$ from the \twoetwov method is plotted together with the synthetic `truth'. It is seen to deviate from the correct value especially around symmetry and in the wings of the mass distribution. If \velpre is never approximated by \velpost in \cref{eq:mpre2e2v,eq:epre2e2v}, the \twoetwov analysis of the synthetic data reproduce the correlations to high precision. This proves that the problem lies in this approximation, but \velpre will of course not be available in an actual experimental situation.

\begin{figure}[tb]
\centering
\includegraphics[width=0.5\textwidth]{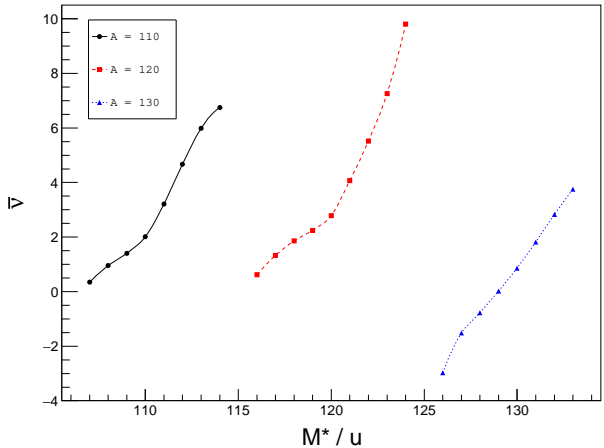}
\caption{\label{fig:acorr}(Color online) The mass-wise positive correlation between the calculated \mpre and \nubar for events with mass number \numlist{110;120;130}, respectively.}
\end{figure}

The assumption of an unchanged fragment velocity, before and after neutron emission, causes each of the velocities to be over- or underestimated with equal chances. This directly leads to an over- and underestimation of the pre-neutron masses in \cref{eq:mpre2e2v}. Since there is no approximation in \cref{eq:mpost2e2v}, $\mpost_1$ and $\mpost_2$ will always be calculated correctly. It is then clear from \cref{eq:nu2e2v} that an overestimation of \mpre directly leads to an overestimation of $\nu$, and vice versa, i.e., there is a positive correlation between the calculated $\mpre$ and $\nu$, as demonstrated in \cref{fig:acorr}, where events of specific mass numbers have been investigated separately. Instead of smoothing the saw tooth shape out, as one could expect from a normal resolution effect, it is exaggerated.

\begin{figure}[tb]
\includegraphics[width=0.5\textwidth]{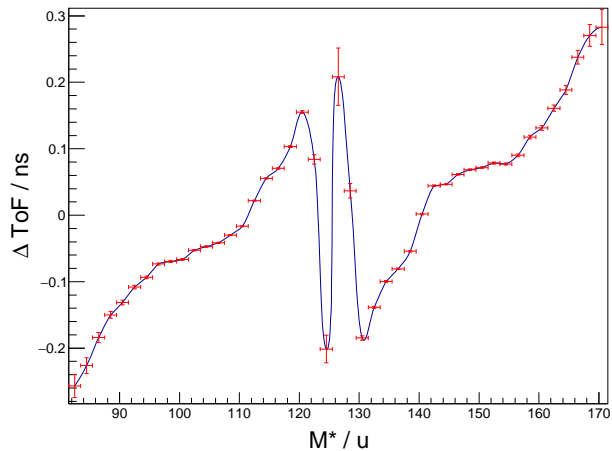}
\caption{\label{fig:deltatof}(Color online) The difference between the ToF calculated by the \twoe method and the true ToF, provided as a function of \mpre. A flight path of \SI{0.5}{\metre} has been assumed.}
\end{figure}

The \twoe method also has a problem, which can affect \twoetwov experiments as well. Examining the \emph{Time-of-Flight} (ToF), calculated from the \twoe-derived \mpost and \epost, reveals discrepancies depicted in \cref{fig:deltatof}. A flight path of \SI{0.5}{\metre} was assumed, corresponding to ToFs in the range of \SIrange{30}{80}{\nano\second}. Since the ToF could not be reproduced, the ratio $\frac{\mpost}{\epost}$ must be incorrect.

Many \twoetwov experiments rely on the \twoe method for calibration \cite{verdi,nishio,velkovska}. Especially setups that use detectors prone to a significant plasma delay time (PDT), like silicon detectors, are in need of a true ToF derived by the \twoe method, in order to correct for the PDT. In a silicon detector, the PDT varies with the particle mass and energy, but is in the order of a few nanoseconds. This means that the deviations from the true value in \cref{fig:deltatof}, for many masses and energies, account for an error of more than \SI{10}{\percent} in the PDT estimation.

The neutron emission does not only add a resolution effect, it creates incorrect correlation in both the examined methods. We will now walk through how that correlation turns into the distorted shape of $\nubar(\mpre)$ seen in \cref{fig:nubar} through an interplay with the mass yield.

The discrepancies are the largest around symmetry, so let us observe \nubar for a mass slightly lower than the mass of symmetric fission (where $\nubar(\mpre)$ is overestimated in \cref{fig:nubar}). Due to the large yield differences between neighboring masses, the calculated \nubar, will be dominated by events that have overestimated \mpre, i.e., events with a lower true \mpre. Since we concluded that these events also overestimates $\nu$, the \nubar evaluated for this mass will be the average of predominately overestimated neutron multiplicities. Thus, \nubar itself will be overestimated. The same effect is mirrored for masses at the other side of symmetry, and the argument applies equally well on the discrepancies in the wings of the mass distribution, where the magnitude of the mass-yield derivative with respect to \mpre is also large. The end result is the exaggerated saw tooth shape one sees in \cref{fig:nubar}.

We suspect that the effect of this correlation can be found in previous measurements. Although the assumption that the velocities are conserved is central to the \twoetwov method, its implications seem to have gone unnoticed in several measurements. In the $^{239}$Pu($n_\text{th}$,$f$) measurements of Nishio~et~al.~\cite{nishio} the velocity assumption is not given much attention and is not even included in the error estimation. Brinkmann~et~al. comments on the errors introduced by neglecting neutron evaporation as ``possibly serious'' but never encounter the correlation problem since they do not present $\nubar(\mpre)$ in their paper \cite{brinkmann}. M\"uller~et~al. report on a \twoetwov measurement of $^{235}$U($n_\text{th}$,$f$) where the over- and undershoot for $\nubar(\mpre)$ in the symmetry region was interpreted as a ``better mass resolution'' (FIG.~11 and~12 in ref.~\cite{muller}). They attributed a pre-neutron mass uncertainty of \SI{0.325}{\amu} to the neutron emission, but they never made a reference to any correction for the correlation problem presented here, so we must assume that their results were affected by it.

Due to the lower neutron multiplicity of $^{235}$U($n_\text{th}$,$f$) compared to $^{252}$Cf(\textit{sf}), the exaggeration of $\nubar(\mpre)$ is less severe in the M\"uller~et~al. data, but we observed the same trend of over- and undershoots by running our simulation also for the $^{235}$U($n_\text{th}$,$f$) case.

By having established the exact cause and effect of the problem, we can now focus on solving it. We will show that it is possible to correct $\nubar(\mpre)$, by outlining a simple \emph{proof-of-concept} correction procedure that do not rely on any new data, only on the measured (uncorrected) pre-neutron mass distribution and neutron multiplicity. The correction is applied to $\nubar(\mpre)$ as a whole, not event-wise, and can therefore also be used to correct previous measurements.

The algorithm works by deconvoluting binned data. In our tests, a bin width of \SI{0.1}{\amu} has been used, which is much smaller than any feasible experimental mass resolution. The origin of the problem is that an actual pre-neutron mass, \mpret, can be incorrectly assigned a different mass \mprem due to the neutron emission. Therefore, we construct a response matrix $R$ such that
\begin {align}
\label{eq:response}\mprem_i = \sum_j R_{ij} \mpret_j.
\end {align}

The response can be closely modeled by a normal distribution, where the mass dependent width can be estimated by kinematic calculations using the measured $\nubar(\mpre)$. In the case of $^{252}$Cf(\textit{sf}) the standard deviation due to neutron emission was on average \SI{0.8}{\amu}. In addition, each matrix element $R_{ij}$ is weighted by the mass yield $Y(\mpret_j)$.

To express the inverse relationship,
\begin {align}
\label{eq:invresponse}\mpret_j = \sum_i R_{ij} \mprem_i.
\end {align}
it was assumed that the transpose of $R$ could approximate $R^{-1}$, even though $R$ is not necessarily strictly orthogonal.

We calculate a corrected \nubar,
\begin{multline}
\nubart_j = \frac{1}{\mn}\left(\mpret_j-\mpost_j\right) = \\
= \frac{1}{\mn}\sum_i R_{ij} \left(\mprem_i-\left(\mpost_i+\Delta M_{ij}\right)\right),
\end{multline}
where $\Delta M_{ij}$ is just the distance between the bins $i$ and $j$ in mass units, and \cref{eq:invresponse} has been used to rewrite $\mpret_j$. We recognize that $\nubarm_i = (\mprem_i-\mpost_i)/\mn$ and arrive at our final expression:
\begin{align}
\nubart_j &= \sum_i R_{ij} \left(\nubarm_i-\frac{\Delta M_{ij}}{\mn}\right).
\end{align}

\begin{figure}[tb]
\centering
\includegraphics[width=0.5\textwidth]{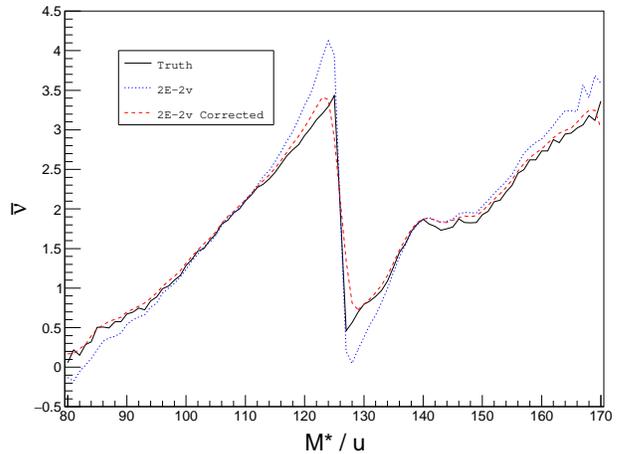}
\caption{\label{fig:corrcf}(Color online) The corrected $\nubar(\mpre)$ from $^{252}$Cf(\textit{sf}) shows signs of smearing, but reproduces the synthetic data well apart from that. The `true' $\nubar(\mpre)$ and the uncorrected $\nubar(\mpre)$ are shown as references. }
\end{figure}

\Cref{fig:corrcf} shows the resulting corrected $\nubar(\mpre)$. The synthetic `true' $\nubar(\mpre)$ is reproduced remarkably well, considering the somewhat crude correction method. The method also conserves the total average neutron multiplicity. Unfortunately, a smearing effect is present, leaving us unable to resolve all structures in the synthetic $\nubar(\mpre)$.

\begin{figure}[tb]
\centering
\includegraphics[width=0.5\textwidth]{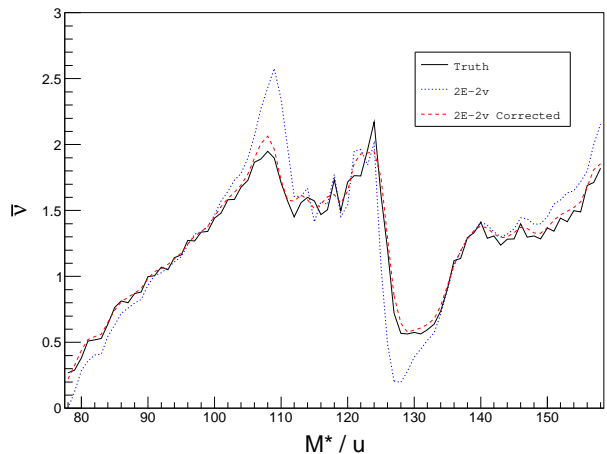}
\caption{\label{fig:corru}(Color online) The correction method behaves the same way applied to $\nubar(\mpre)$ from $^{235}$U($n_\text{th}$,$f$) as when it is applied to $\nubar(\mpre)$ from $^{252}$Cf(\textit{sf}). The `true' $\nubar(\mpre)$ and the uncorrected $\nubar(\mpre)$ are shown as references.}
\end{figure}

In \cref{fig:corru} we show the result of testing the correction on $^{235}$U($n_\text{th}$,$f$) instead of $^{252}$Cf(\textit{sf}). The synthetic \nubar generated by the \textsc{gef} code shows much more structure than one observes in a real experiment on this reaction. However, this makes it an even better test to see how much of the structure in the synthetic data our suggested correction method reproduces.

It has been known from the start that the \twoe method has limited resolution due to the neutron emission. But it is not only a resolution problem. A pure \twoe measurement deduces at least eight quantities, even though only two energies are measured. That is the great success of the method, but it also leaves many degrees of freedom to be incorporated in calibrations or substituted by previously measured data. Reproducing known quantities and correlations is no guarantee that \emph{every} possible correlation deduced by the \twoe method is correct. Caution is advised whenever a new correlation is derived and interpreted. The same analysis used for the experimental data should be tested also with synthetic data, and the dependence of, e.g., the input data and neutron emission should be thoroughly investigated. Only then one is able to decide, whether the quality of the data meets the requirements for the purpose one has in mind.

One such purpose was mentioned previously. Using a \twoe analysis to determine the PDT calibration for a \twoetwov setup, might risk to fail. The deduced ToF deviates systematically from the true value, and neither the \mpre nor the \epost functional dependence are correctly reproduced. The errors would be less significant if a longer flight path was used but that would diminish the solid angle coverage, which is usually already small. Luckily, there are other ways for \twoetwov setups to determine the PDT. Correcting for the PDT is not easy, but examples in literature exist where the PDT corrections did not rely on the \twoe method \cite{brinkmann,muller}.

Measuring \nubar in a straight forward and independent way is one of the highlights of the \twoetwov method. Inability to reproduce the correct shape of $\nubar(\mpre)$ would have made the whole method much less interesting. We found that the intrinsic problem of the \twoetwov method, due to the approximation $\velpre\approx\velpost$, is correctable.

Considering how much effort is currently put into developing the \twoetwov method on various locations around the world, it is of uttermost importance that the consequences of neutron emission gets the attention they deserve, and that the nuclear physics community keep improving on the \twoetwov method itself.

\begin{acknowledgments}
This work was supported by the European Commission within the Seventh Framework Programme through Fission-2013-CHANDA (project no.605203).
\end{acknowledgments}

\bibliographystyle{apsrev}
\bibliography{2e_and_2e2v.bib}

\begin{thebibliography}{16}
\expandafter\ifx\csname natexlab\endcsname\relax\def\natexlab#1{#1}\fi
\expandafter\ifx\csname bibnamefont\endcsname\relax
  \def\bibnamefont#1{#1}\fi
\expandafter\ifx\csname bibfnamefont\endcsname\relax
  \def\bibfnamefont#1{#1}\fi
\expandafter\ifx\csname citenamefont\endcsname\relax
  \def\citenamefont#1{#1}\fi
\expandafter\ifx\csname url\endcsname\relax
  \def\url#1{\texttt{#1}}\fi
\expandafter\ifx\csname urlprefix\endcsname\relax\def\urlprefix{URL }\fi
\providecommand{\bibinfo}[2]{#2}
\providecommand{\eprint}[2][]{\url{#2}}

\bibitem[{\citenamefont{Budtz-Jørgensen
  et~al.}(1987)\citenamefont{Budtz-Jørgensen, Knitter, Straede, Hambsch, and
  Vogt}}]{2e}
\bibinfo{author}{\bibfnamefont{C.}~\bibnamefont{Budtz-Jørgensen}},
  \bibinfo{author}{\bibfnamefont{H.-H.} \bibnamefont{Knitter}},
  \bibinfo{author}{\bibfnamefont{C.}~\bibnamefont{Straede}},
  \bibinfo{author}{\bibfnamefont{F.-J.} \bibnamefont{Hambsch}},
  \bibnamefont{and} \bibinfo{author}{\bibfnamefont{R.}~\bibnamefont{Vogt}},
  \bibinfo{journal}{Nucl. Instrum. Meth. A} \textbf{\bibinfo{volume}{258}},
  \bibinfo{pages}{209 } (\bibinfo{year}{1987}), ISSN \bibinfo{issn}{0168-9002}.

\bibitem[{\citenamefont{Moll et~al.}(1975)\citenamefont{Moll, Schrader,
  Siegert, Asghar, Bocquet, Bailleul, Gautheron, Greif, Crawford, Chauvin
  et~al.}}]{lohengrin}
\bibinfo{author}{\bibfnamefont{E.}~\bibnamefont{Moll}},
  \bibinfo{author}{\bibfnamefont{H.}~\bibnamefont{Schrader}},
  \bibinfo{author}{\bibfnamefont{G.}~\bibnamefont{Siegert}},
  \bibinfo{author}{\bibfnamefont{M.}~\bibnamefont{Asghar}},
  \bibinfo{author}{\bibfnamefont{J.}~\bibnamefont{Bocquet}},
  \bibinfo{author}{\bibfnamefont{G.}~\bibnamefont{Bailleul}},
  \bibinfo{author}{\bibfnamefont{J.}~\bibnamefont{Gautheron}},
  \bibinfo{author}{\bibfnamefont{J.}~\bibnamefont{Greif}},
  \bibinfo{author}{\bibfnamefont{G.}~\bibnamefont{Crawford}},
  \bibinfo{author}{\bibfnamefont{C.}~\bibnamefont{Chauvin}},
  \bibnamefont{et~al.}, \bibinfo{journal}{Nucl. Instrum. Meth.}
  \textbf{\bibinfo{volume}{123}}, \bibinfo{pages}{615 } (\bibinfo{year}{1975}),
  ISSN \bibinfo{issn}{0029-554X}.

\bibitem[{\citenamefont{Oed et~al.}(1984)\citenamefont{Oed, Geltenbort,
  Brissot, Gönnenwein, Perrin, Aker, and Engelhardt}}]{oed}
\bibinfo{author}{\bibfnamefont{A.}~\bibnamefont{Oed}},
  \bibinfo{author}{\bibfnamefont{P.}~\bibnamefont{Geltenbort}},
  \bibinfo{author}{\bibfnamefont{R.}~\bibnamefont{Brissot}},
  \bibinfo{author}{\bibfnamefont{F.}~\bibnamefont{Gönnenwein}},
  \bibinfo{author}{\bibfnamefont{P.}~\bibnamefont{Perrin}},
  \bibinfo{author}{\bibfnamefont{E.}~\bibnamefont{Aker}}, \bibnamefont{and}
  \bibinfo{author}{\bibfnamefont{D.}~\bibnamefont{Engelhardt}},
  \bibinfo{journal}{Nucl. Instrum. Meth.} \textbf{\bibinfo{volume}{219}},
  \bibinfo{pages}{569 } (\bibinfo{year}{1984}).

\bibitem[{\citenamefont{Jansson et~al.}(2017)\citenamefont{Jansson, Fr\'egeau,
  Al-Adili, G\"o\"ok, Gustavsson, Hambsch, Oberstedt, and Pomp}}]{verdi}
\bibinfo{author}{\bibfnamefont{K.}~\bibnamefont{Jansson}},
  \bibinfo{author}{\bibfnamefont{M.~O.} \bibnamefont{Fr\'egeau}},
  \bibinfo{author}{\bibfnamefont{A.}~\bibnamefont{Al-Adili}},
  \bibinfo{author}{\bibfnamefont{A.}~\bibnamefont{G\"o\"ok}},
  \bibinfo{author}{\bibfnamefont{C.}~\bibnamefont{Gustavsson}},
  \bibinfo{author}{\bibfnamefont{F.-J.} \bibnamefont{Hambsch}},
  \bibinfo{author}{\bibfnamefont{S.}~\bibnamefont{Oberstedt}},
  \bibnamefont{and} \bibinfo{author}{\bibfnamefont{S.}~\bibnamefont{Pomp}},
  \bibinfo{journal}{Epj. Web. Conf.} \textbf{\bibinfo{volume}{146}},
  \bibinfo{pages}{04016} (\bibinfo{year}{2017}).

\bibitem[{\citenamefont{Meierbachtol et~al.}(2015)\citenamefont{Meierbachtol,
  Tovesson, Shields, Arnold, Blakeley, Bredeweg, Devlin, Hecht, Heffern,
  Jorgenson et~al.}}]{spider}
\bibinfo{author}{\bibfnamefont{K.}~\bibnamefont{Meierbachtol}},
  \bibinfo{author}{\bibfnamefont{F.}~\bibnamefont{Tovesson}},
  \bibinfo{author}{\bibfnamefont{D.}~\bibnamefont{Shields}},
  \bibinfo{author}{\bibfnamefont{C.}~\bibnamefont{Arnold}},
  \bibinfo{author}{\bibfnamefont{R.}~\bibnamefont{Blakeley}},
  \bibinfo{author}{\bibfnamefont{T.}~\bibnamefont{Bredeweg}},
  \bibinfo{author}{\bibfnamefont{M.}~\bibnamefont{Devlin}},
  \bibinfo{author}{\bibfnamefont{A.}~\bibnamefont{Hecht}},
  \bibinfo{author}{\bibfnamefont{L.}~\bibnamefont{Heffern}},
  \bibinfo{author}{\bibfnamefont{J.}~\bibnamefont{Jorgenson}},
  \bibnamefont{et~al.}, \bibinfo{journal}{Nuclear Instruments and Methods in
  Physics Research Section A: Accelerators, Spectrometers, Detectors and
  Associated Equipment} \textbf{\bibinfo{volume}{788}}, \bibinfo{pages}{59 }
  (\bibinfo{year}{2015}), ISSN \bibinfo{issn}{0168-9002}.

\bibitem[{\citenamefont{Dor\'e et~al.}(2014)\citenamefont{Dor\'e, Farget,
  Lecolley, Lehaut, Materna, Pancin, Panebianco, and Papaevangelou}}]{falstaff}
\bibinfo{author}{\bibfnamefont{D.}~\bibnamefont{Dor\'e}},
  \bibinfo{author}{\bibfnamefont{F.}~\bibnamefont{Farget}},
  \bibinfo{author}{\bibfnamefont{F.-R.} \bibnamefont{Lecolley}},
  \bibinfo{author}{\bibfnamefont{G.}~\bibnamefont{Lehaut}},
  \bibinfo{author}{\bibfnamefont{T.}~\bibnamefont{Materna}},
  \bibinfo{author}{\bibfnamefont{J.}~\bibnamefont{Pancin}},
  \bibinfo{author}{\bibfnamefont{S.}~\bibnamefont{Panebianco}},
  \bibnamefont{and}
  \bibinfo{author}{\bibfnamefont{T.}~\bibnamefont{Papaevangelou}},
  \bibinfo{journal}{Nucl. Data Sheets} \textbf{\bibinfo{volume}{119}},
  \bibinfo{pages}{346 } (\bibinfo{year}{2014}), ISSN \bibinfo{issn}{0090-3752}.

\bibitem[{\citenamefont{Matarranz et~al.}(2013)\citenamefont{Matarranz,
  Tsekhanovich, Smith, Dare, Murray, Pollitt, Soldner, Koster, and
  Biswas}}]{steff}
\bibinfo{author}{\bibfnamefont{J.}~\bibnamefont{Matarranz}},
  \bibinfo{author}{\bibfnamefont{I.}~\bibnamefont{Tsekhanovich}},
  \bibinfo{author}{\bibfnamefont{A.}~\bibnamefont{Smith}},
  \bibinfo{author}{\bibfnamefont{J.}~\bibnamefont{Dare}},
  \bibinfo{author}{\bibfnamefont{L.}~\bibnamefont{Murray}},
  \bibinfo{author}{\bibfnamefont{A.}~\bibnamefont{Pollitt}},
  \bibinfo{author}{\bibfnamefont{T.}~\bibnamefont{Soldner}},
  \bibinfo{author}{\bibfnamefont{U.}~\bibnamefont{Koster}}, \bibnamefont{and}
  \bibinfo{author}{\bibfnamefont{D.}~\bibnamefont{Biswas}},
  \bibinfo{journal}{Physics Procedia} \textbf{\bibinfo{volume}{47}},
  \bibinfo{pages}{76 } (\bibinfo{year}{2013}), ISSN \bibinfo{issn}{1875-3892},
  \bibinfo{note}{scientific Workshop on Nuclear Fission Dynamics and the
  Emission of Prompt Neutrons and Gamma Rays, Biarritz, France, 28-30 November
  2012}.

\bibitem[{\citenamefont{Schmidt et~al.}(2016)\citenamefont{Schmidt, Jurado,
  Amouroux, and Schmitt}}]{gef}
\bibinfo{author}{\bibfnamefont{K.-H.} \bibnamefont{Schmidt}},
  \bibinfo{author}{\bibfnamefont{B.}~\bibnamefont{Jurado}},
  \bibinfo{author}{\bibfnamefont{C.}~\bibnamefont{Amouroux}}, \bibnamefont{and}
  \bibinfo{author}{\bibfnamefont{C.}~\bibnamefont{Schmitt}},
  \bibinfo{journal}{Nucl. Data Sheets} \textbf{\bibinfo{volume}{131}},
  \bibinfo{pages}{107 } (\bibinfo{year}{2016}), \bibinfo{note}{special Issue on
  Nuclear Reaction Data}.

\bibitem[{\citenamefont{Audi et~al.}(2012)\citenamefont{Audi, Wang, Wapstra,
  Kondev, MacCormick, Xu, and Pfeiffer}}]{mass12-1}
\bibinfo{author}{\bibfnamefont{G.}~\bibnamefont{Audi}},
  \bibinfo{author}{\bibfnamefont{M.}~\bibnamefont{Wang}},
  \bibinfo{author}{\bibfnamefont{A.}~\bibnamefont{Wapstra}},
  \bibinfo{author}{\bibfnamefont{F.}~\bibnamefont{Kondev}},
  \bibinfo{author}{\bibfnamefont{M.}~\bibnamefont{MacCormick}},
  \bibinfo{author}{\bibfnamefont{X.}~\bibnamefont{Xu}}, \bibnamefont{and}
  \bibinfo{author}{\bibfnamefont{B.}~\bibnamefont{Pfeiffer}},
  \bibinfo{journal}{Chinese Phys. C} \textbf{\bibinfo{volume}{36}},
  \bibinfo{pages}{1287} (\bibinfo{year}{2012}).

\bibitem[{\citenamefont{Wang et~al.}(2012)\citenamefont{Wang, Audi, Wapstra,
  Kondev, MacCormick, Xu, and Pfeiffer}}]{mass12-2}
\bibinfo{author}{\bibfnamefont{M.}~\bibnamefont{Wang}},
  \bibinfo{author}{\bibfnamefont{G.}~\bibnamefont{Audi}},
  \bibinfo{author}{\bibfnamefont{A.}~\bibnamefont{Wapstra}},
  \bibinfo{author}{\bibfnamefont{F.}~\bibnamefont{Kondev}},
  \bibinfo{author}{\bibfnamefont{M.}~\bibnamefont{MacCormick}},
  \bibinfo{author}{\bibfnamefont{X.}~\bibnamefont{Xu}}, \bibnamefont{and}
  \bibinfo{author}{\bibfnamefont{B.}~\bibnamefont{Pfeiffer}},
  \bibinfo{journal}{Chinese Phys. C} \textbf{\bibinfo{volume}{36}},
  \bibinfo{pages}{1603} (\bibinfo{year}{2012}).

\bibitem[{\citenamefont{Brun and Rademakers}(1996)}]{root}
\bibinfo{author}{\bibfnamefont{R.}~\bibnamefont{Brun}} \bibnamefont{and}
  \bibinfo{author}{\bibfnamefont{F.}~\bibnamefont{Rademakers}},
  \bibinfo{journal}{Nucl. Instrum. Meth. A} \textbf{\bibinfo{volume}{389}},
  \bibinfo{pages}{81} (\bibinfo{year}{1996}).

\bibitem[{\citenamefont{Stein}(1957)}]{stein}
\bibinfo{author}{\bibfnamefont{W.~E.} \bibnamefont{Stein}},
  \bibinfo{journal}{Phys. Rev.} \textbf{\bibinfo{volume}{108}},
  \bibinfo{pages}{94} (\bibinfo{year}{1957}).

\bibitem[{\citenamefont{Nishio et~al.}(1995)\citenamefont{Nishio, Nakagome,
  Kanno, and Kimura}}]{nishio}
\bibinfo{author}{\bibfnamefont{K.}~\bibnamefont{Nishio}},
  \bibinfo{author}{\bibfnamefont{Y.}~\bibnamefont{Nakagome}},
  \bibinfo{author}{\bibfnamefont{I.}~\bibnamefont{Kanno}}, \bibnamefont{and}
  \bibinfo{author}{\bibfnamefont{I.}~\bibnamefont{Kimura}},
  \bibinfo{journal}{J. Nucl. Sci. Technol.} \textbf{\bibinfo{volume}{32}},
  \bibinfo{pages}{404} (\bibinfo{year}{1995}).

\bibitem[{\citenamefont{Velkovska and McGrath}(1999)}]{velkovska}
\bibinfo{author}{\bibfnamefont{J.}~\bibnamefont{Velkovska}} \bibnamefont{and}
  \bibinfo{author}{\bibfnamefont{R.}~\bibnamefont{McGrath}},
  \bibinfo{journal}{Nuclear Instruments and Methods in Physics Research Section
  A: Accelerators, Spectrometers, Detectors and Associated Equipment}
  \textbf{\bibinfo{volume}{430}}, \bibinfo{pages}{507 } (\bibinfo{year}{1999}).

\bibitem[{\citenamefont{Brinkmann et~al.}(1989)\citenamefont{Brinkmann,
  Kiesewetter, Baumann, Freiesleben, Lütke-Stetzkamp, Paul, Schwanke, and
  Sohlbach}}]{brinkmann}
\bibinfo{author}{\bibfnamefont{K.-T.} \bibnamefont{Brinkmann}},
  \bibinfo{author}{\bibfnamefont{J.}~\bibnamefont{Kiesewetter}},
  \bibinfo{author}{\bibfnamefont{F.}~\bibnamefont{Baumann}},
  \bibinfo{author}{\bibfnamefont{H.}~\bibnamefont{Freiesleben}},
  \bibinfo{author}{\bibfnamefont{H.}~\bibnamefont{Lütke-Stetzkamp}},
  \bibinfo{author}{\bibfnamefont{H.}~\bibnamefont{Paul}},
  \bibinfo{author}{\bibfnamefont{H.}~\bibnamefont{Schwanke}}, \bibnamefont{and}
  \bibinfo{author}{\bibfnamefont{H.}~\bibnamefont{Sohlbach}},
  \bibinfo{journal}{Nucl. Instrum. Meth. A} \textbf{\bibinfo{volume}{276}},
  \bibinfo{pages}{557 } (\bibinfo{year}{1989}), ISSN \bibinfo{issn}{0168-9002}.

\bibitem[{\citenamefont{M\"uller et~al.}(1984)\citenamefont{M\"uller, Naqvi,
  K\"appeler, and Dickmann}}]{muller}
\bibinfo{author}{\bibfnamefont{R.}~\bibnamefont{M\"uller}},
  \bibinfo{author}{\bibfnamefont{A.~A.} \bibnamefont{Naqvi}},
  \bibinfo{author}{\bibfnamefont{F.}~\bibnamefont{K\"appeler}},
  \bibnamefont{and} \bibinfo{author}{\bibfnamefont{F.}~\bibnamefont{Dickmann}},
  \bibinfo{journal}{Phys. Rev. C} \textbf{\bibinfo{volume}{29}},
  \bibinfo{pages}{885} (\bibinfo{year}{1984}).

\end{thebibliography}
\end{document}